\documentstyle[doublespacing,referee]{mn}

\input epsf

\def\apj{ApJ}

\def\aap{A\&\hskip-1pt A}

\def\mnras{MNRAS}
\def\pasp{PASP}
\def\araa{ARA\&\hskip-1pt A}

\def\lesssim{\mathrel{\hbox{\rlap{\hbox{\lower4pt\hbox{$\sim$}}}\hbox{$<$}}}}
\def\gtrsim{\mathrel{\hbox{\rlap{\hbox{\lower4pt\hbox{$\sim$}}}\hbox{$>$}}}}

\newcommand{\vvec}  {\mbox{\boldmath $v$}}

\def\eqalign#1{\null\,\vcenter{\openup\jot
        \ialign{\strut\hfil$\displaystyle{##}$&$
        \displaystyle{{}##}$\hfil \crcr#1\crcr}}\,}
\title[Direct Lens Imaging]
       {Direct Lens Imaging of Galactic Bulge Microlensing Events}

\author[Han \& Chang]
{
Cheongho Han$^\dagger$ \& Heon-Young Chang$^\ddagger$\thanks{e-mails: 
cheongho@astroph.chungbuk.ac.kr (CH); hyc@ns.kias.re.kr (HYC)}\\
${}^\dagger$Department of Physics, Institute for Basic Science Research, 
Chungbuk National University, Chongju 361-763, Korea\\
${}^\ddagger$Korea Institute for Advanced Study, 
207-43 Cheongryangri-dong Dongdaemun-gu, Seoul 130-012, Korea
}

\begin{document}
\maketitle
\label{firstpage}
\vspace{-\abovedisplayskip}
\begin{abstract}
Recently, from the {\it Hubble Space Telescope} (HST) images of one of 
the Large Magellanic Cloud (LMC) events taken 6.3 years after the original 
lensing measurement, Alcock et al.\ were able to directly image the lens.
Although the first resolved lens was identified for an LMC event, much 
more numerous lenses are expected to be resolved for Galactic bulge events.
In this paper, we estimate the fraction of Galactic bulge events whose
lenses can be directly imaged under the assumption that all bulge events
are caused by normal stars.  For this determination, we compute the
distribution of lens proper motions of the currently detected Galactic
bulge events based on standard models of the geometrical and kinematical
distributions of lenses and their mass function.  We then apply realistic
criteria for lens resolution, and the result is presented as a function
of the time elapsed after an original lensing measurement, $\Delta t$.
If followup observations are performed by using an instrument with a
resolving power of $\theta_{\rm PSF}=0''\hskip-2pt .1$, which corresponds
to that of HST equipped with the new Advanced Camera for Surveys, we
estimate that lenses can be resolved for $\sim 3\%$ and $22\%$ of disk-bulge
events and for $\sim 0.3\%$ and $6\%$ of bulge self-lensing events after
$\Delta t=10$ and 20 years, respectively.  The fraction increases
substantially with the increase of the resolving power.  If the instrument
has a resolution of $\theta_{\rm PSF}=0''\hskip-2pt .05$, which can be
achieved by the {\it Next Generation Space Telescope}, we estimate that
lenses can be resolved for $\sim 22\%$ and $45\%$ of disk-bulge events and
for $\sim 6\%$ and $23\%$ of bulge self-lensing events after $\Delta t=10$
and 20 years, respectively.

\end{abstract}


\section{Introduction}

Following the proposal of Paczy\'nski (1986), experiments to search for
lensing-induced light variations of stars (microlensing events) located
in the Galactic bulge and the Magellanic Clouds have been or are being
conducted by several groups (MACHO: Alcock et al.\ 1993; EROS: Aubourg
et al.\ 1993; OGLE: Udalski et al. 1993; MOA: Bond et al.\ 2001; DUO:
Alard \& Guibert 1997).  These experiments have successfully detected a
large number of events ($\sim 1,000$), most of which are detected towards
the Galactic bulge.

Despite a large number of event detections, the nature of the lenses is
still poorly known.  This is because the Einstein ring radius crossing
time $t_{\rm E}$ (Einstein timescale), which is the only observable
providing information about the physical parameters of the lens (lens
parameters), results from a combination of the lens parameters, i.e.\
\begin{equation}  
t_{\rm E} = {r_{\rm E}\over v};\qquad
r_{\rm E} = \left[
{4GM\over c^2}{D_{\rm OL}(1-D_{\rm OL})\over D_{\rm OS}}
\right]^{1/2},
\end{equation}
where $r_{\rm E}$ is the Einstein ring radius, $M$ is the lens mass, 
$v$ is the lens-source transverse speed, and $D_{\rm OL}$ and 
$D_{\rm OS}$ are the distances to the lens and the source from the 
observer, respectively.  Under this circumstance, the only approach 
one could pursue would be identifying the major lens population by 
statistically determining the lens mass function based on the observed 
timescale distribution.  However, this approach requires {\it a prior} 
knowledge about the geometrical distribution of the lens, the lens 
kinematics, and the functional form of the lens mass function, which 
are all poorly known.  In addition, even if all lensing objects were 
of the same mass, they would give rise to a broad range of timescale.  
As a result, it is difficult to identify the major lens population 
from this approach (Mao \& Paczy\'nski 1996; Gould 2001).

Recently, from the {\it Hubble Space Telescope} (HST) images of one of
the Large Magellanic Cloud (LMC) events (MACHO LMC-5) taken 6.3 years
after the original lens measurement, Alcock et al.\ (2001) were able 
to resolve the lens from the lensed source star.  By directly imaging 
the lens, they could identify that  the event was caused by a nearby 
low mass star located in the Galactic disk.

Besides the identification of the lens as a normal star, direct lens
imaging is of scientific importance due to following reasons.  First,
by directly and accurately measuring the lens proper motion with respect
to the source, $\mu$, one can better constrain the physical parameters
of the individual lenses.  The previous method to determine $\mu$ is
based on the analysis of the lensing light curves of events affected
by the finite source effect, such as source-transit single lens events
and caustic-crossing binary lens events (Gould 1994; Witt \& Mao 1994;
Nemiroff \& Wickramasinghe 1994).  By analyzing the part of the light 
curves near the source transit or the caustic crossing of these events, 
one can measure the source star angular radius normalized by the angular 
Einstein ring radius $\theta_{\rm E}$, i.e.\ $\rho_\star=\theta_\star/
\theta_{\rm E}$, where $\theta_\star$ is the angular source star radius.  
Then, the lens proper motion is determined by
\begin{equation}
\mu = {\theta_{\rm E}\over t_{\rm E}} = 
      {\theta_\star/\rho_\star \over t_{\rm E}}.
\end{equation}
For the proper motion determination by using this method, however, one
should know the source star angular radius, which can only be deduced
from an uncertain color-surface brightness relation.  As a result, the
proper motions determined in this way suffer from large uncertainties.
By contrast, if the lens is resolved, the proper motion can be directly
and thus accurately measured from the observed image.  Measuring the 
proper motion is equivalent to measuring the angular Einstein ring 
radius because $\theta_{\rm E}=\mu t_{\rm E}$, where the event timescale 
is determined from the light curve.  While $t_{\rm E}$ depends on three 
lens parameters of $M$, $D_{\rm OL}$, and $v$, $\theta_{\rm E}$ does not 
depend on $v$, and thus the lens mass can be better constrained.  Second, 
if the lens is resolved for an event where the lens-source relative 
parallax, $\pi_{\rm rel} = {\rm AU}/(D_{\rm OL}^{-1}-D_{\rm OS}^{-1})$, 
was previously measured during the lensing magnification, one can 
completely break the lens parameter degeneracy and the lens mass is 
uniquely determined by
\begin{equation}
M = {\mu^2 t_{\rm E}^2\over \kappa \pi_{\rm rel}},
\end{equation}
where $\kappa = 4G/(c^2{\rm AU})\sim 8.144\ {\rm mas}/M_\odot$ (Gould
2001).  Third, if the source of an event was resolved via either a source 
transit or a caustic crossing and thus $\rho_\star$ was precisely measured, 
one can determine the angular source star radius by reversing the process 
of the classical method of the proper motion determination, i.e.\ 
$\theta_\star=\mu\hskip2pt t_{\rm E} \hskip2pt \rho_\star$.  By measuring 
$\theta_\star$, one can determine the effective temperature of the source 
star, which is important for the accurate construction of stellar 
atmosphere models (e.g., Alonso et al.\ 2000).

Although the first directly imaged lens was identified for an LMC event, 
much more numerous direct lens identifications are expected if high 
resolution followup observations are performed for events detected 
towards the bulge.  There are several reasons for this expectation.  
First, compared to the total number of LMC events, which is $\sim 20$, 
there are an overwhelmingly large number of bulge events.  Second, 
while the majority of LMC events are suspected to be caused by dark 
(or very faint) objects, most bulge events are supposed to be caused 
by normal stars, for which imaging is possible.  Third, an important 
fraction of lenses responsible for bulge events are believed to be 
located in the Galactic disk with moderate distances, and thus more 
likely to be imaged due to their tendency of being bright and having 
large proper motions.

The goal of this work is to estimate the fraction of Galactic bulge 
events whose lenses can be directly imaged.  For this estimation, we 
first compute the expected distribution of the lens-source proper 
motions of the currently detected Galactic bulge events based on 
standard models of the geometrical and kinematical distributions of 
lenses and their mass function (\S\ 2).  We then apply realistic 
detection criteria for lens resolution and the result is presented as 
a function of the time elapsed after the original lensing measurement,
$\Delta t$ (\S\ 3).  Based on the result in \S\ 3, we discuss some of 
the observational aspects of the future high resolution followup lensing 
observations aimed for direct lens imaging (\S\ 4).  We summarize the 
result and conclude in \S\ 5.

\section{Proper Motions}
The first requirement for direct lens imaging is that the lens should 
have a large relative proper motion with respect to the source so 
that it can be widely separated from the lensed source star within 
a reasonable amount of $\Delta t$.  In this section, we, therefore, 
estimate  the expected distribution of proper motions of Galactic 
bulge events.

With the models of the lens mass function, $\phi(M)$, the matter 
density distributions of the lens and source stars along the line of 
sight towards the Galactic bulge field, $\rho(D_{\rm OL})$ and 
$\rho(D_{\rm OS})$, and the kinematical distribution of lens-source 
transverse velocities, $f(\vvec)$, the distribution of lens proper 
motions of Galactic bulge events is computed by
\begin{equation}
\eqalign{
\Gamma (\mu) \propto &  \int_0^\infty \hskip-8pt dD_{\rm OS}\ \rho(D_{\rm OS}) 
                        \int_0^{D_{\rm OS}} \hskip-10pt dD_{\rm OL}\ \rho(D_{\rm OL})\cr
&  \times  \int_0^\infty \int_0^\infty dv_y dv_z\ v\ f(v_y,v_z) \cr
&  \times  \int_0^\infty dM\ \phi(M)\ r_{\rm E}\ \epsilon(t_{\rm E})\ \delta (\mu - v/D_{\rm OL}),\cr
}
\end{equation}
where $v_y$ and $v_z$ are the components of the transverse velocity which
are respectively parallel with and normal to the Galactic disk plane,
$\epsilon(t_{\rm E})$ is the detection efficiency of events as a function
of $t_{\rm E}$, and the notation $\delta(\cdot\cdot\cdot)$ represents the
Dirac delta function.  We note that the factors $v$ and $r_{\rm E}$ are
included in equation (4) to weight the event rate by the transverse speed
and the lensing cross section.

For the Galactic bulge and disk matter density distributions, we adopt 
the models of Dwek et al.\ (1995) and Bahcall (1986), respectively.  
In the bulge model, the bulge has a triaxial shape and the matter 
density is represented by an analytic form of
\begin{equation}
\rho(r_s) \propto \exp (0.5 r_s^2);\ \ 
r_s^4 = [(x'/x_0)^2+(y'/y_0)^2]^2+(z'/z_0)^4,
\end{equation}
where 
$(x_0,y_0,z_0)=(1.58, 0.62, 0.43)$ kpc, and the coordinates $(x',y',z')$ 
represent the axes of the triaxial bar from the longest to the shortest, 
and the longest axis is misaligned with the line of sight toward the 
Galactic center by an angle $\theta=20^\circ$.  The Bahcall disk model 
is expressed by a double exponential form of 
\begin{equation}
\rho(R,z) \propto \exp \left[ -\left( {R-8000\over h_R} + {z\over h_z}
\right)\right],
\end{equation}
where the radial scale length and the vertical scale height are $h_z=325$ 
pc and $h_R=3.5$ kpc, respectively.

For the transverse velocity distribution, we adopt the model of Han \&
Gould (1995).  In this model, the velocity distributions for both disk
and bulge components have a Gaussian form of
\begin{equation}
f(v_i)\propto {\rm exp}\ \left[ -{(v_i-\bar{v}_i)^2\over 2\sigma_i^2}\right],
\qquad
i \in y,\ z.
\end{equation}
The means and the standard deviations of the individual velocity components 
for events with disk lenses and bulge source stars (disk-bulge events) are
\begin{equation}
\eqalign{
 & (\bar{v}_y,\sigma_y)=(220.0,30.0)\ {\rm km\ s}^{-1}, \cr
 & (\bar{v}_z,\sigma_z)=(0.0,20.0) \ {\rm km\ s}^{-1}, \cr
}
\end{equation}
where $\bar{v}_y=220\ {\rm km}\ {\rm s}^{-1}$ corresponds to the rotation 
speed of the Galactic disk and $(\sigma_y, \sigma_z)= (30, 20) \ {\rm km}\ 
{\rm s}^{-1}$ are the adopted velocity dispersions of stars in the solar 
neighborhood.  For events with bulge source stars lensed by another 
foreground bulge stars (bulge self-lensing events), the means and standard 
deviations of the transverse velocity distributions are
\begin{equation}
\eqalign{
 & (\bar{v}_y,\sigma_y)=(0.0,82.5)\ {\rm km\ s}^{-1},  \cr
 & (\bar{v}_z,\sigma_z)=(0.0,66.3) \ {\rm km\ s}^{-1}.\cr
}
\end{equation}
For the barred bulge, the velocity dispersions along the axes of the bar 
are deduced from the tensor virial theorem (Binney \& Tremaine 1987), 
resulting in $(\sigma_{x'},\sigma_{y'},\sigma_{z'})=(113.6, 77.4, 66.3)$ 
${\rm km}\ {\rm s}^{-1}$.  Due to the projection effect caused by the bar 
misalignment, the projected velocity dispersions are computed by
$(\sigma_{y},\sigma_{z})=
([\sigma_{x'}^2\sin^2\theta+\sigma_{y'}^2\cos^2\theta]^{1/2}, \sigma_{z'})=
(82.5, 66.3)\ {\rm km}\ {\rm s}^{-1}$, which correspond to the 
standard deviations in (9).  We note that the adopted velocity model is 
a rough approximation in the sense that it does not include factors 
such as the systematic mean motion of bulge stars discussed by Evans \& 
Belokurov (2002), the figure rotation of the bulge discussed by Blum 
(1995), and the possibility of non-Gaussian nature of disk star velocity 
distribution discussed by Evans \& Collett (1993), which may actually be 
important in the resulting distribution of proper motions.  Since the 
inclusion of the mean motion or the figure rotation of bulge stars
result in high proper motions, we note that the fraction of resolvable 
lenses predicted by the adopted bulge velocity model is the lower limit.

The Einstein timescale distribution of events observed by the MACHO 
group is claimed to be consistent with the distribution from normal 
stars (Alcock et al.\ 2000b).  We, therefore, model the mass function 
of lenses based on the present day main-sequence stars determined by
Kroupa, Tout \& Gilmore (1993).  The adopted mass function has a three 
power-law functional form of
\begin{equation}
\phi(M) \propto
\cases{
M^{-1.3} & for $0.08\ M_\odot \leq M < 0.5\ M_\odot$ \cr
M^{-2.2} & for $0.5\ M_\odot \leq M < 1.0\ M_\odot$ \cr
M^{-4.5} & for $M \geq 1.0\ M_\odot$. \cr
}
\end{equation}
For the detection efficiency, we adopt the latest determination of the 
MACHO experiment (Alcock et al.\ 2000b), whose data were analyzed by 
using the `difference image analysis' method.

\begin{figure}
\epsfysize=12.0cm
\vskip-0.5cm
\centerline{\epsfbox{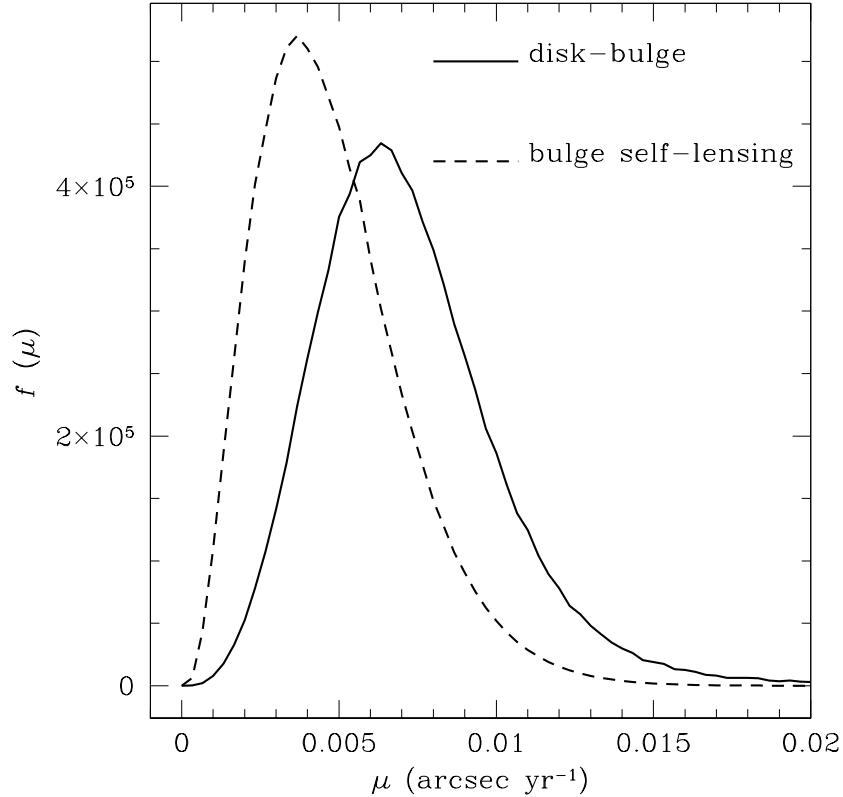}}
\vskip-0.2cm
\caption{
The expected distributions of relative lens-source proper motions for 
Galactic bulge events.  The solid curve is for events with disk lenses 
and bulge source stars (disk-bulge events), while the dashed curve is 
for events with bulge source stars lensed by another foreground bulge 
stars (bulge self-lensing events).  The ordinate of each distribution 
is arbitrarily normalized so that the areas under the two curves match 
together.
}
\end{figure}

\begin{figure}
\epsfysize=12.0cm
\vskip-0.2cm
\centerline{\epsfbox{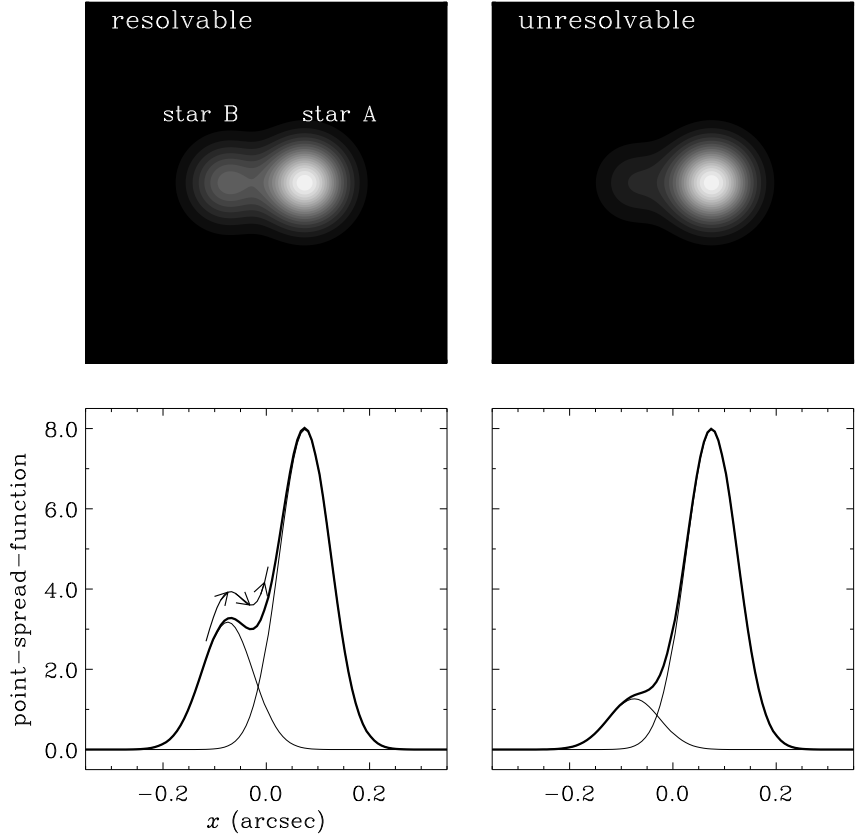}}
\caption{
Criterion for resolving two closely located stars.  The presented in 
each of the upper panels is the combined image of two stars and the 
thick solid curve in the corresponding lower panel is the one-dimensional 
PSF profile of the combined image.  The two stars comprising the 
individual combined images have a common separation of $0''\hskip-2pt .15$, 
but have different magnitude differences of $\Delta m=1.0$ (left image) 
and 2.0 (right image).  The PSF of each star (thin solid curves in each 
of the lower panels) is modeled by a Gaussian, where the standard deviation 
$\sigma$ represents the resolving power of the instrument to be used for 
imaging.  Here we assume $\sigma=0''\hskip-2pt .05$.  We assume that the 
individual stars are resolved if the sign of the combined image's PSF 
profile changes more than twice in the overlapping region between the 
centers of the individual stellar images.  Under this criterion, the two
stars in the left panel is resolved, while the stars in the right panel 
is not resolved.
}
\end{figure}

In Figure 1, we present the obtained distributions of relative
lens-source proper motions for disk-bulge (solid curve) and bulge
self-lensing (dashed curve) events, respectively.  Due to the large 
uncertainties in the geometrical and kinematical distributions of lenses, 
the relative contribution of the disk and bulge lenses to the total event 
rate is very uncertain.  We, therefore, leave the proper motion
distributions of the disk-bulge and bulge self-lensing events separately,
instead of estimating the distribution of total events by arbitrarily 
normalizing the ratio between the two populations of events.  From the 
figure, one finds that, as expected, the average proper motion of events 
caused by disk lenses is larger than that of events caused by bulge lenses.

\section{Further Restriction}

The lens detectability is additionally restricted by the lens brightness.  
This is not only simply because the lens should be brighter than a 
detection limit, but also because the threshold lens-source separation 
for lens detection, $\theta_{\rm th}$, varies depending on the apparent 
lens/source flux ratio.  Therefore, we compute the angular threshold 
as a function of the magnitude difference between the lens and the source, 
$\Delta m=m_{\rm L} -m_{\rm S}$.

For the computation of $\theta_{\rm th}(\Delta m)$, we first model the
point-spread-function (PSF) of a stellar image as a Gaussian, where the
standard deviation $\sigma$ of the PSF characterizes the resolving power
of the instrument to be used for followup lensing observations.  We then
generate the combined image of the lens and the source by normalizing 
such that the volume under each PSF is proportional to the flux of each 
star.  Once the combined image is constructed, we then judge whether the 
lens and the source can be resolved each other.  For this judgment, we 
assume that the lens can be resolved if the sign of the derivative of the 
combined image's one-dimensional PSF profile changes more than two times 
in the overlapping region between the centers of the lens and the source 
images (see Figure 2 for illustration).  Then, the angular threshold is 
defined as the lens-source separation at which the lens is just to be 
resolved from the source star.

In Figure 3, we present the computed angular threshold as a function 
of $\Delta m$.  The two curves correspond to the distributions expected 
when the followup observations are carried out by using two different 
instruments, whose resolving powers are characterized by $\theta_{\rm PSF}
=2\sigma=0''\hskip-2pt .1$ (dashed curve) and $0''\hskip-2pt .05$ (solid 
curve), respectively.  We note that the Advanced Camera for Surveys (ACS) 
recently installed on HST can achieve a resolution of $\theta_{\rm PSF}
\sim 0''\hskip-2pt .1$.  We also note that the Near Infrared Camera 
(NIRCam) of {\it Next Generation Space Telescope} (NGST), which will have 
an aperture of 6 -- 7 m, will be sensitive in the wavelength range of 0.6 
to 5 microns, and thus can achieve $\theta_{\rm PSF}\sim 0''\hskip-2pt .05$.  
From the figure, one finds that as the lens becomes fainter, it becomes 
difficult to resolve the lens from the source star.

\begin{figure}
\epsfysize=12.0cm
\vskip-0.5cm
\centerline{\epsfbox{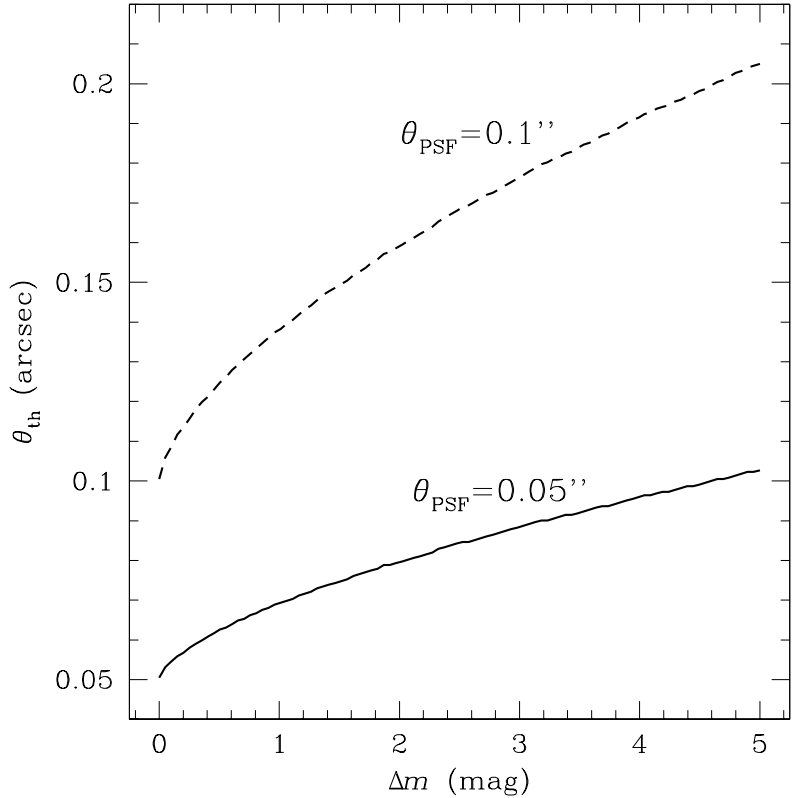}}
\vskip-0.2cm
\caption{
The threshold angular separation for resolving two closely located stars 
as a function of magnitude difference between the two stars.  The angular 
threshold is computed for two instruments whose resolving power is 
characterized by $\theta_{\rm PSF}=0''\hskip-2pt .1$ and 
$\theta_{\rm PSF}=0''\hskip-2pt .05$.
}
\end{figure}

With the computed distribution of $\theta_{\rm th}(\Delta m)$, we 
recalculate the distribution of proper motions of events by using 
equation (4), but in this time only for events with detectable lenses.
To be detected, the lens should meet the condition of $\mu \Delta t 
\geq \theta_{\rm th}(\Delta m)$.  In addition, we also restrict that 
detectable lenses should be brighter than a threshold magnitude of 
$I=22$.  For this computation, we assume that the source star has a 
fixed brightness of $M_I=3.0$, which corresponds to that of a bulge 
clump giant, while the lens brightness is deduced from its mass by 
using the mass-luminosity relation provided by Kroupa, Tout \& Gilmore 
(1993).  Once the absolute magnitudes of the source and the lens are 
set, their apparent magnitudes are computed considering their distances 
from the observer, i.e.\ $D_{\rm OS}$ for the source and $D_{\rm OL}$ 
for the lens.

\begin{figure}
\epsfysize=12.0cm
\vskip-0.5cm
\centerline{\epsfbox{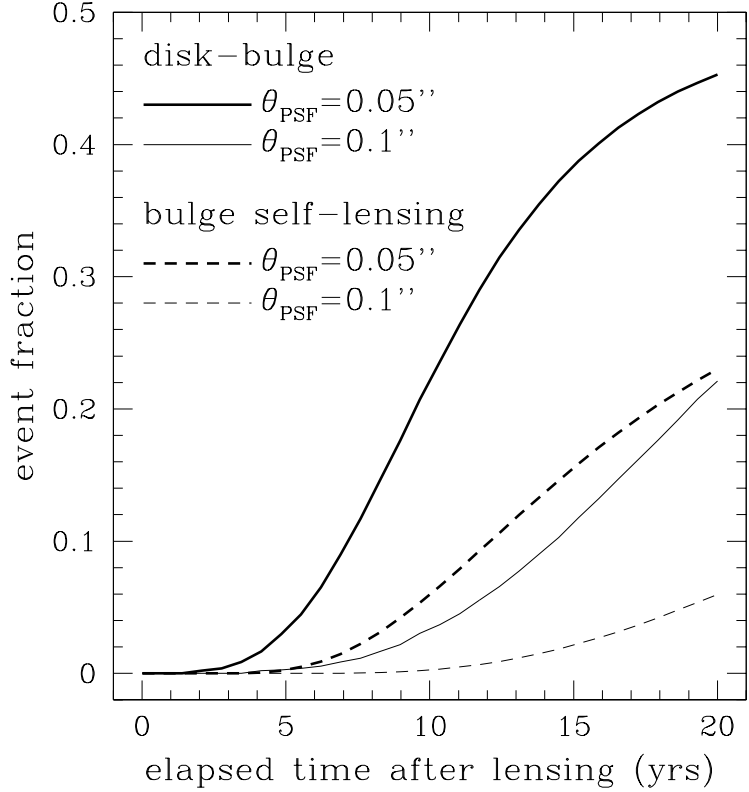}}
\vskip-0.2cm
\caption{
The fractions of events with lens-source separations $\Delta\theta=\mu
\Delta t \geq \Delta\theta_{\rm th}$ as a function of elapsed time after
lensing magnification $\Delta t$. To estimate the fraction the actual
minimum separation is considered due to the difference in magnitude
for a given  $\theta_{\rm PSF}$.  The two pairs of curves drawn by
continuous lines and dashed lines correspond to the fractions obtained
with two different PSF separation of $\Delta\theta_{\rm PSF}=
0''\hskip-2pt .05$ and $0''\hskip-2pt .1$.  The red curves are for
disk-bulge events while the blue curves are for bulge-bulge events.
}
\end{figure}

In Figure 4, we present the finally determined fractions of events with 
detectable lenses as a function of $\Delta t$.  In the figure, the thick 
and thin curves represent the expected fractions when followup observations 
are conducted by using instruments with $\theta_{\rm PSF}=0''\hskip-2pt .05$ 
and $0''\hskip-2pt .1$, respectively.  If the instrument has a resolution 
of $\theta_{\rm PSF}=0''\hskip-2pt .1$, we estimate that lenses can be 
resolved for $\sim 3\%$ and $22\%$ of disk-bulge events and for 
$\sim 0.3\%$ and $6\%$ of bulge self-lensing events after $\Delta t=10$ 
and 20 years, respectively.  The fraction increases substantially with 
the increase of the resolving power.  If followup observations are 
performed by using an instrument with $\theta_{\rm PSF}=0''\hskip-2pt .05$,
we estimate that lenses can be resolved for $\sim 22\%$ and $45\%$ of 
disk-bulge events and for $\sim 6\%$ and $23\%$ of bulge self-lensing 
events after $\Delta t=10$ and 20 years, respectively.

\section{Discussion}

In the previous section, we estimated the fraction of Galactic bulge 
events for which one can directly image lenses from followup observations 
by using high resolution imaging instruments and presented the result as 
a function of a time after original lensing measurements.  In this section, 
based on the result in the previous section we discuss some of the 
observational aspects of the future followup lensing observations aimed 
for direct lens imaging.

First, from the distributions in Fig.\ 4, we find that the proper choice 
of the instrument for the future high resolution followup lensing 
observations will be NGST.  Bulge events has been reported since 1993 
(Udalski et al.\ 1994; Alcock et al.\ 1995a).  Under a rough assumption 
that disk-bulge and bulge self-lensing events equally contribute to the 
total Galactic bulge event rate and considering the life expectancy of 
HST, the fraction of events with resolvable lenses from HST observations 
will be just $\sim 5\%$ even if the followup observations are performed 
at the end stage of HST for the first generation of lensing events.  
However, by using NGST, which is scheduled to be launched in 2009, it 
will be possible to resolve lenses for a significant fraction of events.

Second, given that all lensing events will be unable to be followed up 
due to the limited observation time of NGST, priority of targets should 
be given to events from which one can obtain extra information about the 
lens or the source star other than the identification of the lens as a 
normal star and the measurement of the lens-source proper motion.  As 
mentioned in \S\ 1, one can measure the angular radius of the source star 
of an event, for which the source was previously resolved, and uniquely 
determine the lens mass of an event, for which the lens parallax was 
measured.  Therefore, these events should be at the top of the target list.
In the published literature, we find 10 events, for which the 
source star was well resolved (Alcock et al.\ 1997, 2000a; Albrow et al.\ 
1999, 2000, 2001a,b; Afonso et al.\ 2000; An et al.\ 2002)\footnote{Among 
them, one event was detected towards the Small Magellanic Cloud (MACHO 
98-SMC-1).}, and 11 events, for which parallax effect was measured 
(Alcock et al.\ 1995b; Mao 1999; Soszy\'nski et al.\ 2001; Smith, Mao, 
\& Wo\'zniak 2002).  Another possible high priority targets will be the 
longest events with $t_{\rm E}\gtrsim 70$ days. These events cannot be 
well explained by the standard models of the geometrical and kinematical 
distributions of lenses and their mass function (Han \& Gould 1996).
Although these events comprise a small fraction ($\lesssim 10\%$) of 
the total number of events, they are important because their contribution 
to the total optical depth is important.  Detection of lenses and the 
measured proper motions (or non-detection of lenses) will allow one to 
better constrain the nature of these mysterious events.

\section{Summary and Conclusion}
We have estimated the fraction of Galactic bulge microlensing events 
for which the lenses can be directly imaged from future high resolution 
followup observations by computing the distribution of proper motions 
of the currently detected bulge events and imposing realistic criteria for 
lens resolution.  From this computation, we find that lens identification 
will be possible for a significant fraction of bulge events from followup 
observations using NGST under the assumption that most bulge events 
are caused by normal stars.  Besides identifying lenses as stars, direct 
lens imaging will allow one to accurately determine the lens proper motion,
from which the physical parameters of the individual lenses can be better 
constrained.  If lenses are imaged for events where lens parallaxes were 
measured, the lens parameter degeneracy can be completely broken and 
the lens mass can be uniquely determined.  In addition, high resolution 
followup observations will provide a valuable chance to measure the 
angular radii of remote bulge source stars involved with events for which 
the source was previously resolved via either a source transit or a 
caustic crossing.

We would like to thank A.\ Gould for proving useful comments about the
work.  This work was supported by a grant (R01-1999-00023) of the Korea
Science \& Engineering Foundation (KOSEF).

{}

\end{document}